# Low-Loss THz Waveguide Bragg Grating using a Two-Wire Waveguide and a Paper Grating

Guofeng Yan,[1,3] Andrey Markov,[1] Yasser Chinifooroshan,[2] Saurabh M. Tripathi,[2] Wojtek J. Bock,[2] and Maksim Skorobogatiy[1,*]

[1] *Department of Engineering Physics, Ecole Polytechnique de Montréal, Québec, Canada*
[2] *Département d'informatique et d'ingénierie, Université du Québec en Outaouais, Québec, Canada*
[3] *Centre for Optical and Electromagnetic Research, Zhejiang University, P. R. China*
*Corresponding author: maksim.skorobogatiy@polymtl.ca*

We propose a novel kind of the low-loss THz Waveguide Bragg Grating (TWBG) fabricated using plasmonic two-wire waveguide and a micromachined paper grating for potential applications in THz communications. Two TWBGs were fabricated with different periods and lengths. Transmission spectra of these TWBGs show 17 dB loss and 14 dB loss in the middle of their respective stop bands at 0.637 THz and 0.369 THz. Insertion loss of 1-4 dB in the whole 0.1-0.7 THz region was also measured. Finally, TWBG modal dispersion relation, modal loss and field distributions were studied numerically, and low-loss, high coupling efficiency operation of TWBGs was confirmed.

The terahertz (THz) spectrum has seen significant technological advances over the past few decades, which were mainly driven by applications in sensing [1], industrial characterization [2], and fundamental photonics [3]. Recently, THz communications became an active research topic, driven by the availability of unregulated bandwidth and a promise of much higher transmission rates as compared to microwave communications. Communication applications rely heavily on the availability of high quality signal delivery and signal processing components such as waveguides and waveguide Bragg gratings. However, design of low-loss, high coupling efficiency optical components in THz spectral range proved to be challenging due to high losses and high material dispersion of the materials. Recently, fabrication of the THz fiber Bragg gratings was reported in [4, 5] using laser inscription on plastic subwavelength fibers. However, subwavelength fibers suffer from low operational bandwidth and high dispersion [6], which limits the THz communication bandwidth and the signal transmission length.

In this work, we demonstrate fabrication and optical characterization of a novel kind of low-loss, high coupling efficiency THz Waveguide Bragg Grating using the combination of a low-loss, low-dispersion plasmonic two-wire waveguide and a low-loss micromachined paper grating. We believe that such TWBGs can pave the way for novel high quality signal processing components for demanding THz communication applications.

We start by discussing optical properties of the fundamental mode of a THz Waveguide Bragg Grating. In our calculations we use COMSOL Multiphysics FEM software to solve for the modal complex effective refractive indices and field profiles. The TWBG considered in this work is composed of a paper grating inserted between two copper wires (see Fig. 1(a)). For simplicity of presentation, the paper grating with pitch Λ=225 μm is replaced by the uniform paper layer of the same thickness H=100 μm, while Bragg condition is imposed via simple reflection of the modal dispersion relation at the edge of the first Brillouin zone $k_{BZ} = \pi / \Lambda$ (see Fig. 1(b)). In our simulations we consider that the real part of the paper reflective index is frequency independent and equal to 1.45, while the paper absorption coefficient is frequently dependent $\alpha \left[ cm^{-1} \right] = 41.33 * \upsilon^2 + 2.039$, where operation frequency $\upsilon$ is in THz (see [5] for details).

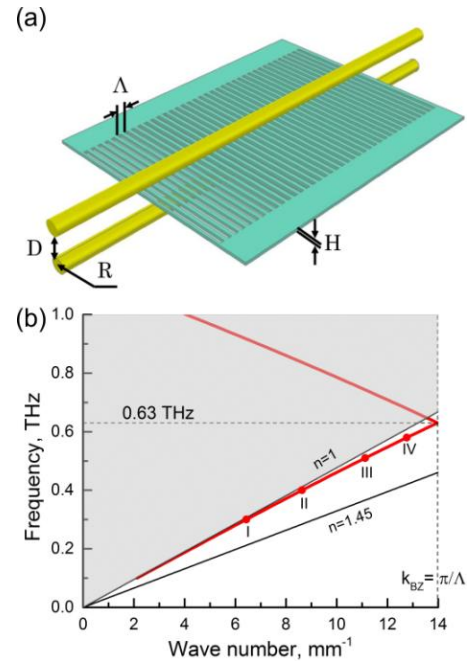

Fig. 1. (a) Schematic of a THz Waveguide Bragg Grating. (b) Approximate band diagram of the fundamental mode of a TWBG.

We note that although a full 3D model of a TWBG can be readily analyzed, we only seek to explain general properties of the fundamental grating mode such as field distribution as a function of frequency and an approximate position of the stop band, for which a simplified model is adequate.

The wire diameter is set to R=250 µm, and the gap between the wires is D=900 µm. The frequency dependent relative permittivity and conductivity of metal are modeled using the Drude formula for copper, which in THz spectral range predicts frequency independent real part of the relative permittivity $\varepsilon_r = -1.7 \cdot 10^5$, and frequency independent conductivity $\sigma = 4.5 \cdot 10^7 \, S/m$.

In Fig. 1 (b) (in red) we present dispersion relation of the TWBG fundamental mode with a stop band expected at 0.63 THz. At frequencies lower than ~0.65 THz the fundamental mode is guided by the total internal reflection. At higher frequencies the modal dispersion relation is reflected into the continuum of radiative states (gray area in the band diagram), which in practice manifests itself in the form of increased transmission loss.

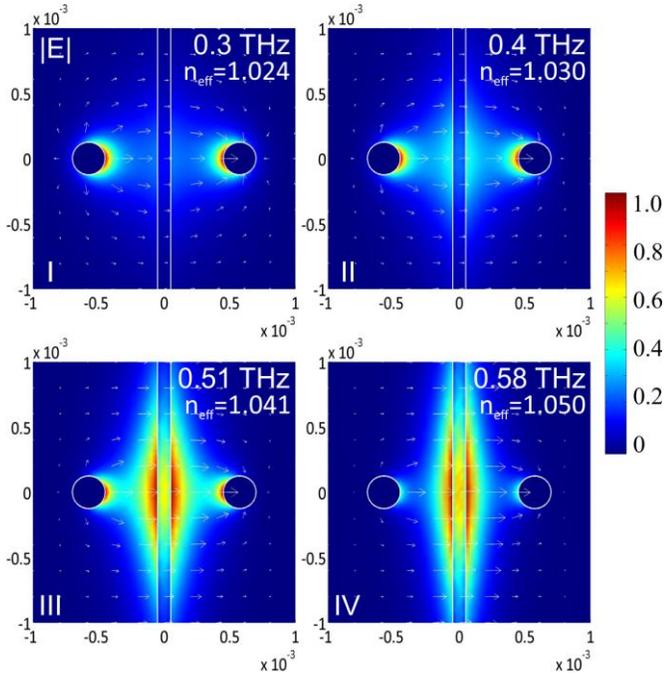

Fig. 2. Longitudinal flux distribution for the fundamental plasmonic mode of a TWBG at various operation frequencies. Arrows show the direction of the corresponding electric fields.

In Fig. 2 we present longitudinal flux distribution and electric vector field distribution for the fundamental plasmonic mode of a TWBG at various operation frequencies. At low frequencies (< 0.45 THz), electric field of the fundamental mode is mostly oriented along the line connecting the two wires. The corresponding energy flux is concentrated in the near vicinity of the wires with very little power found inside of the paper layer. This field distribution is very similar to that of a plasmonic mode of a classic two-wire THz waveguide. As a result, the modal loss is considerably smaller than the bulk absorption loss of paper (see Fig. 3 (a)), while larger than the loss of a corresponding two-wire waveguide without paper. When increasing the operation frequency, one observes stronger presence of the modal field in the lossy paper region, thus leading to rapid increase in the modal losses.

Coupling efficiency into a TWBG (defined as a fraction of the coupled power to the power in the excitation beam) was computed assuming as an excitation source a Gaussian THz beam perfectly centered between the wires with a frequency-dependent beam diameter $d_0 \approx 2.5 \cdot \lambda$. The excitation efficiency of a TWBG is compared with that of a two-wire waveguide in Fig. 3 (b). The presence of a paper layer enables better confinement of the modal power between the wires, thus, leading to a much higher excitation efficiency of the TWBG mode (up to 80%) compared to that of a corresponding classic two-wire waveguide (efficiency up to 40%). Coupling into the fundamental mode of a TWBG is the most efficient in the vicinity of 0.5 THz.

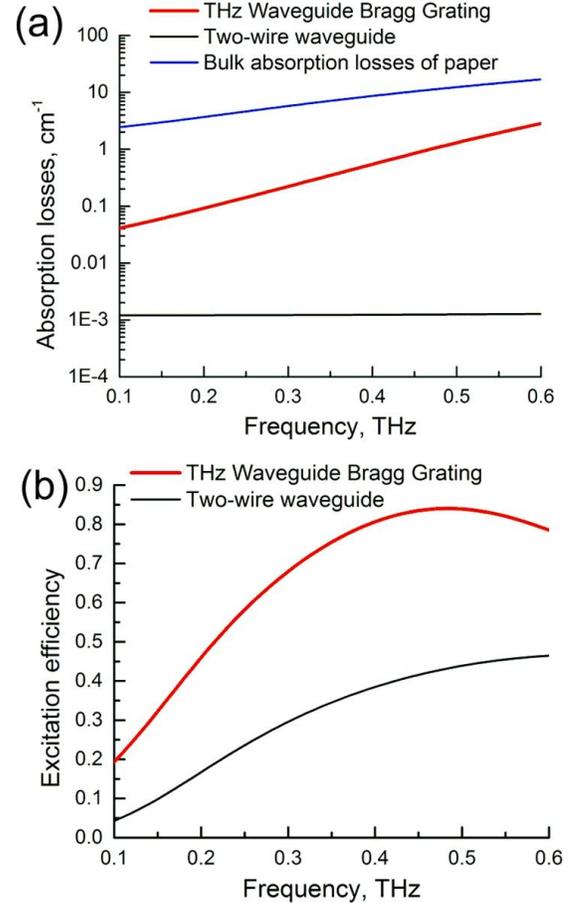

Fig.3. Comparison of (a) absorption losses and (b) excitation efficiencies of the fundamental mode of a TWBG and the fundamental mode of a two-wire waveguide.

In order to optimize THz waveguide Bragg grating performance it is desirable, on one hand, to enhance grating strength by increasing the field presence in the grating region. On the other hand, one has to keep modal loss and modal dispersion at minimum by ensuring that a significant amount of energy is still guided outside of the paper region. In this respect, as seen in Fig 2, operation frequency around 0.5 THz is probably the optimal one as it satisfies both of the optimization criteria. Namely, there is a significant field presence both in the paper grating and in the plasmon mode outside of the grating region. Unfortunately, this frequency is not matched with spectral position of the TWBG stop band. By changing separation between the two wires, and by varying thickness of the grating layer, we believe that it is possible to explore the tradeoff between grating strength and modal loss and realize optimal operation regime in the vicinity of a desired stop band. The question of TWBG optimization is, however, beyond the scope of this paper and we will explore it further in our future work.

Experimentally, the THz Waveguide Bragg Gratings were fabricated by precision laser cutting of slit arrays in a regular 100 μm-thick printing paper of refractive index ~1.45. Class IV Synrad $CO_2$ laser operating at 10.6 μm with average output power of 1.5 W was used for cutting (beam spot size < 100 μm). Two batches of paper gratings are presented in this work, one consists of 90 slits that are ~110 μm-wide with a pitch of 226 μm, and a total grating length ~2.1 cm (see Fig. 4 (a)). The other one consists of 60 slits that are ~190 μm-wide with a pitch of 370 μm (see Fig. 4 (b)). As one can see in Fig. 4 (a) and (b), the borders of the slits are not smooth and charred because of the laser burning the paper. However, these features are deeply subwavelength and the paper itself is low loss in the THz [7], so the resulting grating loss is low.

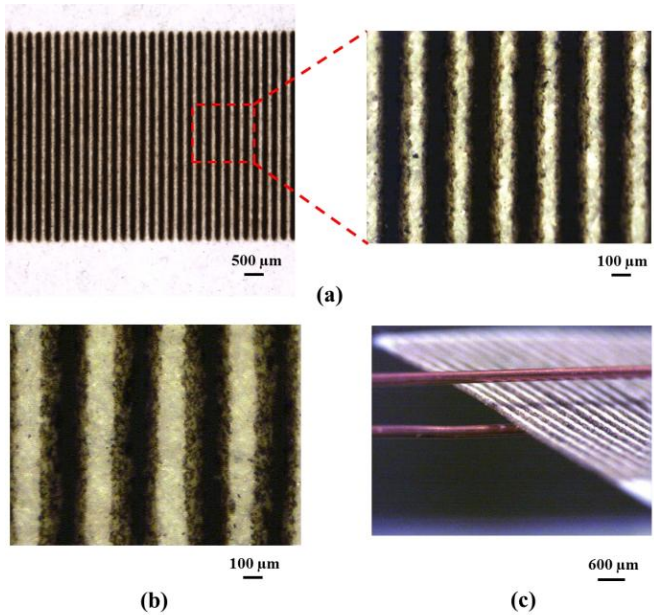

Fig. 4. (a) and (b) Microscopy photos of paper gratings with different periods. (c) The side view of the paper grating inserted between the two copper wires.

The TWBG transmission spectra were recorded using a THz-time domain spectroscopy (TDS) setup modified for the waveguide measurements [6]. In our studies, we used ~600 ps-long scans that define spectral resolution of ~1.5 GHz. In order to increase grating strength the paper grating was inserted between the two copper wires (of length ~5 cm). The separation between the two copper wires was 900 μm and the paper grating was positioned in the middle of the gap under microscope (see Fig.4 (c)). The plasmonic two-wire waveguides are known to be wide-band, low-loss and low-dispersion [8, 9], which is also seen in our reference measurement (Fig.5, middle).

In Fig. 5, we present three experimentally measured traces of the THz electric field. One is a reference THz pulse registered without any waveguide. Second is a THz pulse after transmission through an empty 5 cm-long two-wire waveguide. Third is a THz pulse after transmission through a 5 cm-long two-wire waveguide with a 2 cm-long paper grating inside. As expected, transmission through an empty two-wire waveguide does not change significantly the pulse shape, while transmission through a two-wire waveguide with paper grating shows an enlarged pulse and a ringing effect. This allows us a direct estimate of the waveguide dispersion, which is proportional to the increase in the pulse duration and inversely proportional to the pulse bandwidth and the length of the grating. From Fig. 5 we estimate the dispersion to be less than 1 ps / (cm THz). We believe that modal dispersion can be greatly reduced by TWBG optimization via reduction of the space between the wires and using thinner paper gratings.

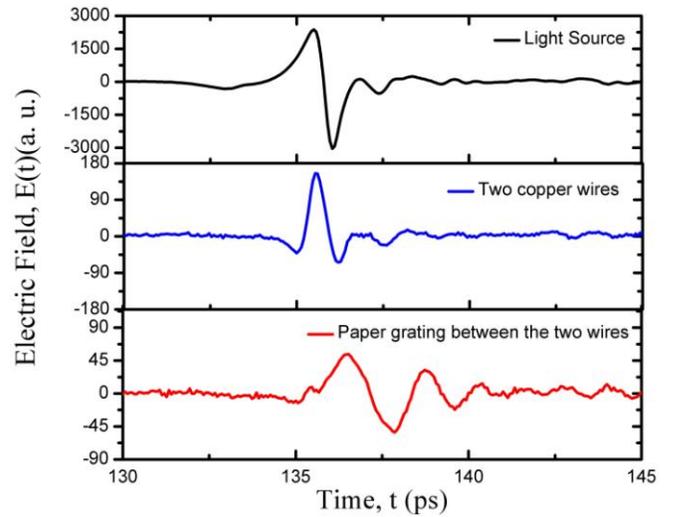

Fig. 5. Experimentally measured electric field traces

Figure 6 presents transmission spectra (in the frequency domain) that correspond to traces shown in Fig. 5. A transmission band from 0.1 THz to 1.5 THz is observed in the transmission spectrum of an empty two copper wire waveguide (black curve). When paper grating is inserted between the two wires, TWBG bandwidth decreases to 0.1 THz - 0.87 THz (green curve) due to additional material losses incurred in the paper grating, which is especially pronounced at higher frequencies. After normalization (Fig. 6, bottom) a 4.6 GHz-wide stop band (Full Width at Half Maximum) at ~637 GHz and ~17 dB transmission loss in the middle of a stop band are observed (blue curve). Using GratingMOD module from RSoft Design Group with geometrical parameters discussed earlier, predicts the value of a stop band center frequency to be ~635 GHz, which is in good agreement with the experimental measurements. At this frequency, the TWBG modal refractive index is ~1.05. We also note that experimental insertion loss of a TWBG is only ~ 4 dB in the vicinity of a stop band, while the insertion loss is much smaller (<1dB) at lower frequencies 0.1~0.5 THz. This is also in accordance with simulation results.

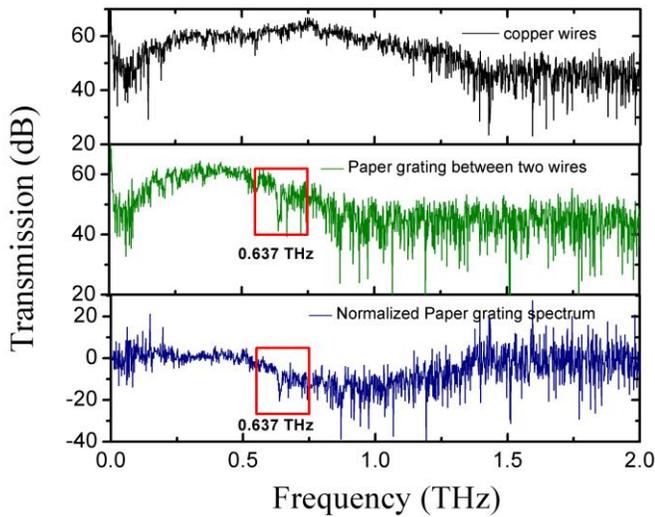

Fig. 6. Transmission spectra for (top) a two-wire waveguide, (middle) a two-wire waveguide with a paper grating with 226 μm grating pitch. Normalized paper grating spectrum for insertion loss calculation (bottom).

By modifying the grating period we can place the stop band anywhere in the THz spectral range. Thus, by using a longer pitch of 370 μm (see Fig. 4 (b)) we have realized another grating with a stop band centered at lower frequency ~0.369 THz. As we can see from the normalized transmission spectrum presented in Fig. 7, the insertion loss varies between 1-4 dB in the 0.2 ~0.5 THz range, while the stop band transmission loss is ~14 dB. Numerically, GratingMOD predicts Bragg frequency of 0.378 THz and modal refractive index ~1.03, which is, again, in good agreement with experimental observations.

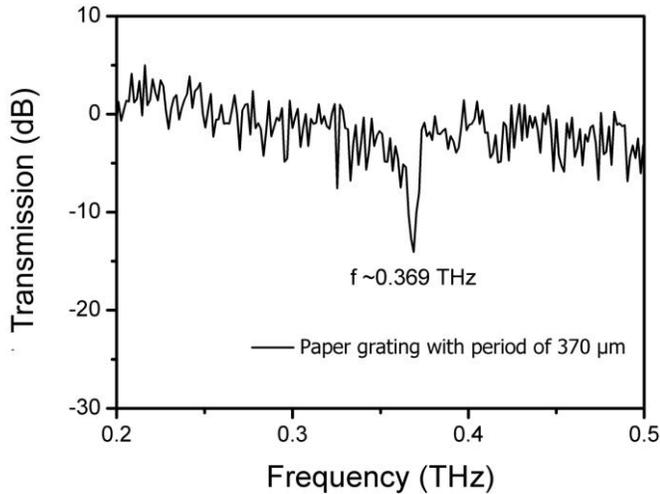

Fig. 7. Normalized transmission spectrum of a TWBG with grating pitch of 370 μm.

In conclusion, a novel low-loss THz Waveguide Bragg Grating was numerically investigated and experimentally demonstrated by using plasmonic two-wire waveguide and a paper grating inserted between the two wires. Transmission spectrum of the TWBG with 90 periods shows ~17 dB transmission loss in the middle of a stop band centered at ~0.63 THz, and an insertion loss of 1- 4 dB in the whole 0.1~0.7 THz region. By adjusting the grating period, position of the TWBG stop band can be changed at will. We demonstrated this by realizing another grating with a stop band centered at ~0.37 THz and having a ~14 dB transmission dip. We believe that presented low-loss, high coupling efficiency grating design can pave the way for the development of novel low-loss signal processing components for demanding THz communication and sensing applications.